# Differential Filtering in a Common Basic Cycle: Multi-Major Trajectories and Structural Bottlenecks in Exact Sciences and Engineering Degrees

Hugo Roger Paz
PhD Professor and Researcher Faculty of Exact Sciences and Technology National University of Tucumán
Email: hpaz@herrera.unt.edu.ar
ORCID: https://orcid.org/0000-0003-1237-7983

## ABSTRACT

Universities often present the Common Basic Cycle (CBC) as a neutral levelling stage shared by several degree programmes. Using twenty years of longitudinal administrative records from a Faculty of Engineering and Exact Sciences, this study tests whether the CBC actually operates as a uniform gateway or as a differential filter across majors. We reconstruct student trajectories for 24,017 entrants, identifying CBC subjects (year level ≤ 1), destination major, time to exit from the CBC, and final outcome (progression to upper cycle, drop-out, or right-censoring). The analysis combines transition matrices, Kaplan–Meier survival curves, stratified Cox models and subject-level logistic models of drop-out after failure, extended with multi-major enrolment data and a pre/post 2006 curriculum reform comparison.

Results show that the CBC functions as a strongly differential filter. Post-reform, the probability of progressing to the upper cycle in the same major ranges from about 0.20 to 0.70 across programmes, while overall drop-out in the CBC exceeds 60%. Early Mathematics modules (introductory calculus and algebra) emerge as structural bottlenecks, combining low pass rates with a two- to three-fold increase in the hazard of leaving the system after failure, with markedly different severity by destination major. Multi-major enrolment, often treated administratively as indecision, is instead associated with lower drop-out, suggesting an adaptive exploration of feasible trajectories. The findings portray the CBC not as a neutral academic foyer, but as a structured sorting device whose impact depends sharply on the targeted degree and on the opportunity to explore alternative majors.

# 1. INTRODUCTION

Common basic cycles and first-year common cores have become a widespread response to the massification of higher education. Rather than relying on competitive entrance examinations, many universities now admit students into a shared set of foundational modules in mathematics, physics, chemistry and introductory disciplinary content, and only later channel them into specific degree programmes. In Argentina, the *Ciclo Básico Común* (CBC) at the University of Buenos Aires, created in 1985, is a paradigmatic example: its implementation required a thorough reorganisation of curricula, new teaching sites and the consolidation of the CBC as a separate academic unit with its own governance and academic staff (Buchbinder, 2014; Universidad de Buenos Aires, n.d.). The CBC model has since inspired local variants in other public institutions, particularly in engineering and exact sciences, where a common first year is often justified as a levelling device that guarantees minimum academic standards while preserving open access.

Empirically, however, first-year curricula in engineering and exact sciences are also the main locus of attrition. Studies using longitudinal data in STEM degrees repeatedly show that a large fraction of departures occurs in the first two years of study, frequently concentrated around mathematically intensive modules and introductory physics (Paura & Arhipova, 2014). These early losses are not merely a matter of academic performance in a single examination. They arise from the interaction between demanding syllabi, compressed assessment schedules, unfamiliar institutional cultures and, frequently, precarious socio-economic conditions. When these elements align unfavourably, students may disengage from the programme or the institution even if they remain formally enrolled for some time.

Survival-type analyses, which explicitly model time to dropout and handle censored trajectories, have begun to clarify this temporal dimension. Applying proportional hazards models to engineering cohorts, Paura and Arhipova (2014) showed that dropout risk is heavily front-loaded, with secondary-school preparation and early academic results exerting strong effects in the first semesters. More recently, leakage-aware data frameworks have emphasised the need to treat student trajectories as path-dependent processes constrained by curriculum structure, institutional shocks and policy changes, rather than as independent observations in static prediction models (Paz, 2025). Within such frameworks, first-year subjects function not only as "difficult courses" but as structural nodes where different pathways diverge.

Despite this progress, two important blind spots remain. First, the CBC and similar common cores are often treated as homogeneous experiences. Quantitative analyses typically aggregate at the level of broad fields (for example, "engineering" versus "non-engineering") or at the level of individual subjects, implicitly assuming that a given module exerts a similar selective pressure regardless of the student's intended degree. This assumption is particularly strong in faculties where the same CBC subjects are shared by degrees with markedly different mathematical intensity, professional identities and labour-market horizons. It is plausible, for instance, that the same introductory calculus course is experienced as routine by students targeting long-cycle engineering degrees and as a high-stakes hurdle by those aiming at shorter or more applied programmes in computing or applied sciences.

Secondly, there is limited empirical work on **multi-major enrolment** within common basic cycles. Regulations often allow – and even encourage – simultaneous registration in more than one degree during the CBC, under the rationale that students should be able to explore alternatives before committing to a specific programme. In practice, multi-major trajectories are usually collapsed into a single "home" major for analytical purposes. This obscures whether students use the CBC as a deliberate exploration space – moving, for example, from a shorter programming degree into a longer engineering degree once they have "tested" their performance – or whether multiple registrations are a symptom of indecision and increased risk.

A third gap concerns the impact of curriculum reforms on the filtering role of the CBC over time. In many Latin-American engineering faculties, substantial reforms to first-year structures were introduced in the mid-2000s, including the extension or reorganisation of common cores, the tightening of prerequisite structures and the redefinition of progression rules (Buchbinder, 2014). While policy documents often frame these reforms as improvements in academic quality and retention, there is relatively little longitudinal evidence on how they modify survival patterns and transitions across majors. It remains unclear whether reformed CBC structures operate as more efficient levelling mechanisms or simply intensify early selection without improving eventual graduation.

This paper addresses these gaps using longitudinal administrative data from a Faculty of Engineering and Exact Sciences at a large public university. Building on a leakage-aware data architecture for student analytics (Paz, 2025), we reconstruct the CBC histories of several cohorts spanning two decades, identifying CBC subjects, time to exit from the CBC and outcomes (progression to the upper cycle in the same major, progression to a different major, dropout, or right-censoring). Crucially, we distinguish between **initially declared major**, **destination major** and the **number of majors held during the CBC**, and we explicitly separate cohorts entering before and after the implementation of a major CBC reform in 2005–2006.

The study is guided by four research questions:

1. **RQ1.** Does the Common Basic Cycle operate as a uniform or differential filter across degree programmes in engineering and exact sciences, particularly after the 2005–2006 curriculum reform?

2. **RQ2.** How do subject-level pass rates and failure-related dropout risks vary by destination major?

3. **RQ3.** Is multi-major enrolment during the CBC associated with greater persistence, strategic reorientation, or increased dropout risk?

4. **RQ4.** To what extent do identified bottleneck and sentinel subjects account for differential trajectories across majors?

Methodologically, we combine transition matrices, Kaplan–Meier survival curves and stratified Cox models with subject-level logistic models that estimate the odds of dropping out after failing specific CBC subjects. Multi-major information is integrated into a student-level dataset that captures entry conditions, CBC performance, number and sequence of majors, and outcomes. By treating the CBC as a **shared but structurally differentiated** layer – rather than as a neutral gateway – we aim to show that the same formal common core can embody markedly different filtering profiles across degree programmes and cohorts. These findings have direct implications for the design of first-year structures, the regulation of multi-major enrolment and the identification of structural bottlenecks in engineering and exact-science curricula.

## 2. METHODS

### 2.1. Institutional context and data sources

The study was conducted in a Faculty of Engineering and Exact Sciences at a large public university in Argentina. The faculty offers long-cycle engineering degrees (for example, Civil, Mechanical, Chemical, Electronic, Electrical, Industrial and Computer Engineering), shorter technical degrees (for example, University Programmer) and bachelor-level programmes in exact sciences (for example, Mathematics).

We used the full institutional academic records stored in the central information system, including all enrolments, examination registrations, examination results, subject approvals and degree registrations. The observation window covers more than two decades of cohorts entering through the Common Basic Cycle (CBC). The raw dataset comprises 996,806 academic events linked to 30,885 student–major combinations, from which 24,017 unique students with CBC activity were identified.

## 2.2. Cohort definition and inclusion criteria

The analytic population consists of all students who:

1. Registered in at least one subject classified as part of the CBC (see Section 2.3); and
2. Had at least one degree programme registered in the faculty; and
3. Entered the faculty within the global observation window (entry year between 1980 and 2019).

Students with incomplete identifiers or inconsistent dates (for example, negative time between CBC entry and last recorded activity) were retained for descriptive checks but excluded from survival models through listwise deletion.

For analyses involving the curriculum reform, cohorts were split into two groups:

- **Pre-reform cohorts:** students with CBC entry year earlier than 2006.
- **Post-reform cohorts:** students with CBC entry year of 2006 or later, corresponding to the implementation of a revised CBC structure with modified subject composition and progression rules.

**Table 1. Cohort characteristics and outcomes by major before and after the 2006 reform.**

| Major Destination | N (Pre-2006) | CBC Dropout (Pre) | N (Post-2006) | CBC Dropout (Post) |
|---|---|---|---|---|
| Ingeniería Industrial | 726 | 44.1% | 1387 | 52.3% |
| Ingeniería en Computación | 1904 | 64.3% | 1143 | 75.8% |
| Ingeniería Química | 1083 | 60.5% | 1105 | 58.4% |
| Ingeniería Mecánica | 1302 | 58.1% | 1059 | 65.8% |
| Ingeniería Civil * | 2050 | 57.6% | **752** | 74.6% |
| Ingeniería Electrónica | 1588 | 33.9% | 687 | 67.2% |
| Ingeniería Biomédica | 470 | 36.8% | 670 | 66.6% |

| Major Destination | N (Pre-2006) | CBC Dropout (Pre) | N (Post-2006) | CBC Dropout (Post) |
|---|---|---|---|---|
| Tec. Univ. Tecnol. Azucarera | 310 | 52.6% | 430 | 41.4% |
| Programador Universitario | 1899 | 80.5% | 319 | 76.5% |
| Ingeniería Eléctrica | 289 | 37.7% | 240 | 60.0% |

**2.3. Operationalisation of the Common Basic Cycle**

Rather than relying on programme labels alone, the CBC was operationalised structurally as the subset of subjects that satisfy all of the following conditions:

1. They are classified at **year level 1** in the official curriculum structure for at least one degree;
2. They are shared by two or more degree programmes in engineering or exact sciences; and
3. They correspond to foundational areas (for example, Mathematics, Physics, Chemistry, introductory disciplinary modules, general academic skills).

A reference table of CBC subjects (*CBC_subjects_ref*) was constructed from the institutional curriculum catalogues and manually checked. For each subject we recorded:

- Subject code and name;
- Field group (Mathematics, Physics, Chemistry, Introductory Engineering, Computing, Language, Other);
- List of degree programmes for which the subject forms part of the official CBC route.

**2.4. Construction of the student-level dataset**

All analyses are based on a leakage-aware student-level dataset (*CBC_student_level_extended*) that summarises entry conditions, CBC performance, major trajectories and outcomes for each student. The construction followed three main steps.

**2.4.1. Initial declared major, destination major and multi-major trajectories**

For each student we identified:

- **Initial declared major.** The first degree programme registered at or before the CBC start date. If multiple majors were declared on the same date, the major that later became the destination major was preferred; otherwise, the major with the lowest identifier was selected. A binary flag records cases where the initial major had to be imputed using the earliest registration within 12 months of CBC entry.

- **Destination major.** The degree in which the student first passed a subject classified as part of the upper cycle (year level ≥ 2) after the CBC. If the student never passed any upper-cycle subject, the declared major at the time of their last CBC activity was used as the provisional destination major.

- **Number of majors during the CBC.** The number of distinct degree programmes for which the student had active registration between CBC entry and CBC exit (see Section 2.4.3). This variable is denoted *n_majors_during_CBC*.

Using these elements, a categorical pattern variable was derived:

- *no_switch* (initial and destination major coincide);

- *switch_within_engineering* (both majors are engineering degrees but differ);

- *switch_programmer_to_engineering* (initial degree is University Programmer, destination degree is an engineering programme);

- *switch_other* (all remaining combinations).

**2.4.2. CBC performance indicators**

CBC performance was summarised using the following indicators:

- Number of CBC subjects attempted, passed and failed;

- Proportion of CBC subjects passed (CBC pass rate at student level);

- Mean CBC grade (when numeric marks were available);

- Number of CBC subjects passed within the first 12 months after CBC entry (*cbc_subjects_passed_first_year*);

- Binary indicators for ever having failed specific key subjects (for example, introductory calculus, algebra, introductory computing, physics), separately for "first attempt" and "any attempt".

All CBC performance variables were computed without using information from upper-cycle subjects or later semesters, in order to avoid temporal leakage.

### 2.4.3. Time-to-event variables and outcomes

For each student, the start of exposure to the CBC was defined as the date of the first enrolment in any CBC subject (*cbc_start_date*). The exit date from the CBC (*cbc_end_event_date*) was defined in a mutually exclusive way as:

1. The date of the first passed upper-cycle subject in the destination major (progression to upper cycle in the same major);
2. The date of the first passed upper-cycle subject in a different major (progression to upper cycle in another major);
3. The date of the last CBC activity (enrolment or examination event) followed by at least three years of inactivity in all subjects (CBC dropout);
4. The date of the last academic event when the observation window ends before conditions (1)–(3) are met (right-censored).

The time-to-event variable (*time_to_event*) was measured in months between CBC entry and CBC exit, using year and month information from the academic calendar.

The outcome variable *cbc_outcome* assumed four values:

- *upper_destination_major* (progression to upper cycle in the destination major);
- *upper_other_major* (progression to upper cycle in a different major);
- *dropout_cbc* (cessation of all academic activity for at least three years after CBC);
- *censored* (no event before the end of the observation window).

For modelling purposes, a numeric version (*cbc_outcome_code*) was also created (1 = upper_destination_major; 2 = dropout_cbc; 3 = upper_other_major; 0 = censored).

### 2.5. Identification of bottleneck and sentinel subjects

To identify structural bottlenecks, we first computed, for each CBC subject:

- Total number of enrolments;
- Number and proportion of passes at the level of the whole faculty;
- Mean time to first pass among those who eventually passed.

A **bottleneck score** was then defined as:

$$\text{bottleneck score}_j = (1 - \text{pass rate}_j) \times \text{HR}_{j,\text{dropout}},$$

where HR$_{j,\text{dropout}}$ is the hazard ratio for CBC dropout associated with ever failing subject $j$, estimated from Cox models adjusted for CBC performance and entry cohort (see Section 2.6). Subjects with high bottleneck scores combine low pass rates with a strong association between failure and subsequent dropout.

To identify **sentinel subjects** whose failure predicts dropout in a major-specific way, logistic regression models were fitted at subject–major level (Hosmer et al., 2013). For each key CBC subject, a model was estimated with:

- Outcome: dropout during the CBC (versus any form of progression or censoring);
- Predictors: ever having failed the subject (binary), destination major, and their interaction;
- Controls: CBC pass rate, CBC mean grade, and entry year.

A subject–major pair was labelled sentinel when all of the following conditions were satisfied:

1. Prevalence of failure ≥ 0.10 in the major;
2. Adjusted odds ratio for dropout after failure ≥ 2.0;
3. p-value ≤ .01 for the failure effect in that major.

Language electives and subjects with very low enrolments were excluded from this sentinel analysis.

**Table 2. Sentinel Subjects: Combinations of major and subject with the highest associated risk of drop-out upon failure.**

| Major | Sentinel Subject (Materia) | Prevalence of Failure | Risk Multiplier (Odds Ratio) |
|---|---|---|---|
| Tec. Univ. en Iluminación | Representación Gráfica | 80% | 7.87 |
| Tec. Univ. en Iluminación | Laboratorio de Comp. I | 75% | 6.71 |
| Programador Universitario | Cálculo I | 52% | 3.96 |
| Lic. en Informática | Laboratorio I | 96% | > 3.50 |
| Ingeniería en Computación | Elementos de Computación | 55% | 3.44 |

## 2.6. Analytical strategy

The analysis proceeds in four stages.

First, we describe CBC trajectories using **transition matrices** from CBC entry to three mutually exclusive outcomes at a fixed time horizon of three years: progression to the upper cycle in the destination major, progression to the upper cycle in another major, and CBC dropout. Transition probabilities are computed separately by destination major and by entry period (pre- versus post-2006), yielding an aggregate picture of how the CBC filters students across programmes and over time.

Secondly, we use **survival analysis** to model time to exit from the CBC. Kaplan–Meier curves are estimated for each destination major, treating any exit (progression or dropout) as an event and censoring students whose trajectories remain active at the end of the observation window (Singer & Willett, 2003). To quantify covariate effects, we fit Cox proportional hazards models stratified by destination major, with predictors including CBC pass rate, number of CBC subjects passed in the first year and entry cohort (Therneau & Grambsch, 2000). These models provide hazard ratios that summarise how early CBC performance accelerates or delays exit from the CBC, while allowing each major to have its own baseline hazard function.

Thirdly, we examine **multi-major trajectories** using the extended student-level dataset. We summarise the distribution of *n_majors_during_CBC* and estimate outcome probabilities (progression in the same major, progression in another major, dropout) by initial major and by number of majors. Transition tables from initial to destination major allow us to quantify, for example, how often students who start as University Programmers later become Computer Engineering students, and whether such shifts differ between pre- and post-reform cohorts.

Finally, we relate subject-level and major-level patterns by combining bottleneck scores and sentinel flags with the transition and survival analyses. Subject-level indicators reveal which CBC modules contribute most strongly to early attrition, while major-level outcomes indicate how these bottlenecks map onto specific degree programmes.

## 2.7. Ethical considerations

All analyses were conducted on de-identified administrative records provided by the university under a data-use agreement that restricts access to authorised researchers. Student identifiers were replaced by pseudonymous codes before analysis, and results are reported only in aggregate form. No attempt was made to re-identify individuals or to report cell counts that could compromise anonymity. The study complies with institutional guidelines for the ethical use of academic records in research.

## 3. RESULTS

### 3.1. Overall outcomes in the Common Basic Cycle

Across the full observation window, 24,017 students entered the Common Basic Cycle (CBC) with at least one degree registered in the Faculty of Engineering and Exact Sciences. Within a three-year horizon from CBC entry, the dominant outcome was **dropout during the CBC**: slightly over 60% of students ceased all academic activity without ever passing an upper-cycle subject. Only around 35–40% progressed to the upper cycle of their destination major, and transitions to the upper cycle of a different major were rare.

These patterns are not stable over time. When cohorts are split into pre- and post-reform groups (entry before versus from 2006 onwards), progression probabilities in most degrees decline by 10–20 percentage points, with corresponding increases in CBC dropout. In post-reform cohorts, the CBC functions as a markedly more selective layer, with higher early attrition and only modest gains in progression (Figure 1).

**Figure 1.** Student outcomes in the Common Basic Cycle by initial declared major. Stacked bars show the proportion of students who dropped out (teal), progressed to the upper cycle in the same major (yellow), or remained censored (purple) at the end of the observation window. Programmes are ordered by dropout rate. *Note: "Unknown" refers to students whose initial major could not be reliably determined from registration records.*

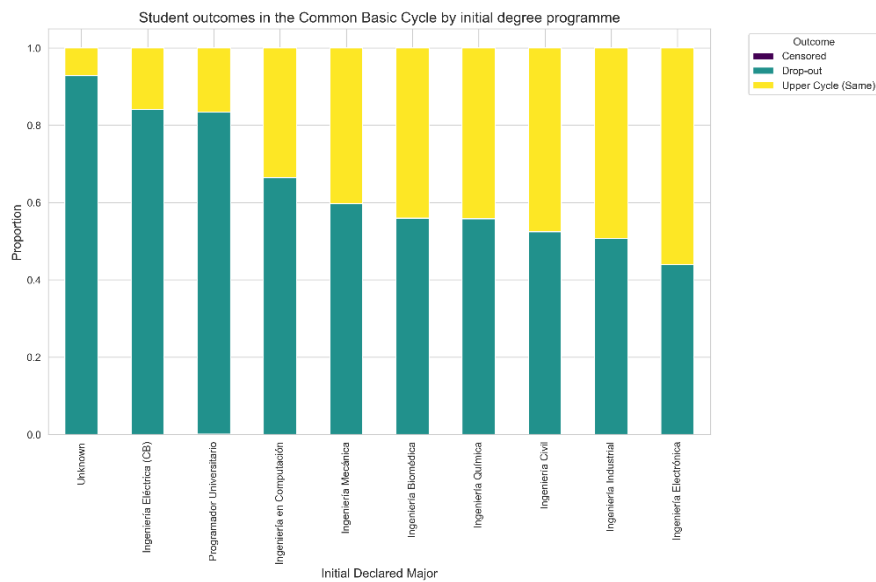

### 3.2. Differential filtering across degree programmes

The aggregate statistics conceal substantial variation by degree. Even within engineering and exact sciences, the CBC does **not** act as a uniform gateway.

Instead, it behaves as a **differential filter** whose intensity depends on the destination programme.

In post-reform cohorts, the probability of reaching the upper cycle in the *same* degree ranges approximately:

- from about 0.45 to 0.55 in programmes such as Industrial, Electronic and, in earlier cohorts, Civil Engineering;
- to around 0.30–0.40 in Mechanical and Chemical Engineering;
- and down to roughly 0.20–0.25 for shorter technical and applied degrees such as University Programmer and certain exact-science degrees.

Conversely, CBC dropout probabilities range from around 0.45 in the most permeable engineering degrees to 0.75–0.80 in shorter and more applied programmes. For some degrees with relatively small cohorts but demanding first-year structures, the CBC operates almost as a pure expulsion device: only a small minority of entrants reach the upper cycle, and most leave the system within the first two or three years.

Transitions from the CBC to the upper cycle of a **different** degree are negligible in aggregate. This indicates that, in this institutional context, the CBC does **not** function as a substantial reorientation mechanism between majors; it primarily sorts students into persistence versus dropout within their declared trajectories. Table 3 summarises the distribution of multi-major enrolment during the CBC and its association with outcomes: students who held two or more majors exhibit markedly lower dropout rates (47% versus 64%) and higher progression to the upper cycle compared to those who remained in a single programme throughout the CBC.

**Table 3.** Distribution of multi-major enrolment during the Common Basic Cycle and associated outcomes. Students holding more majors during the CBC show progressively lower dropout rates and higher progression to the upper cycle.

| n_majors_during_CBC | n_students | proportion_students | proportion_dropout_cbc | proportion_progress_upper_same_major |
|---|---:|---:|---:|---:|
| 1 | 21185 | 0.88 | 0.64 | 0.36 |
| 2 | 2518 | 0.10 | 0.47 | 0.53 |
| 3 | 263 | 0.01 | 0.46 | 0.54 |
| 4 | 46 | 0.00 | 0.37 | 0.63 |
| 5 | 5 | 0.00 | 0.40 | 0.60 |

## 3.3. Time to exit from the Common Basic Cycle

Survival curves reinforce this picture. Kaplan–Meier estimates show that the probability of remaining in the CBC falls steeply during the first 24–30 months after entry. The median time to exit (by either progression or dropout) is around **two years**, with interquartile ranges of approximately 18–30 months.

When stratified by destination major, the shapes of the survival curves are similar, but their positions differ. Students in more permeable engineering degrees tend to exit the CBC slightly earlier and are more likely to do so via progression rather than dropout. In contrast, students in shorter technical and applied degrees often leave the CBC rapidly through dropout, with relatively fewer surviving long enough to accumulate sufficient passes for progression. There is little evidence of students remaining in the CBC for extended periods without an event; the CBC behaves as a **transitory but high-stakes layer**, not as a long-term holding zone.

Stratified Cox models confirm that early CBC performance is strongly associated with time to exit. Controlling for entry cohort, each additional CBC subject passed in the first year substantially increases the hazard of exiting the CBC, while higher CBC grade-point averages are also associated with faster exit. Figure 2 displays the Kaplan-Meier survival curves stratified by destination major, showing that exit probabilities converge rapidly across programmes during the first 24 months but diverge thereafter, with Electronic Engineering students exhibiting the longest retention times and most programmes reaching near-complete exit by month 200.

**Figure 2.** Kaplan-Meier survival estimates for retention in the CBC.

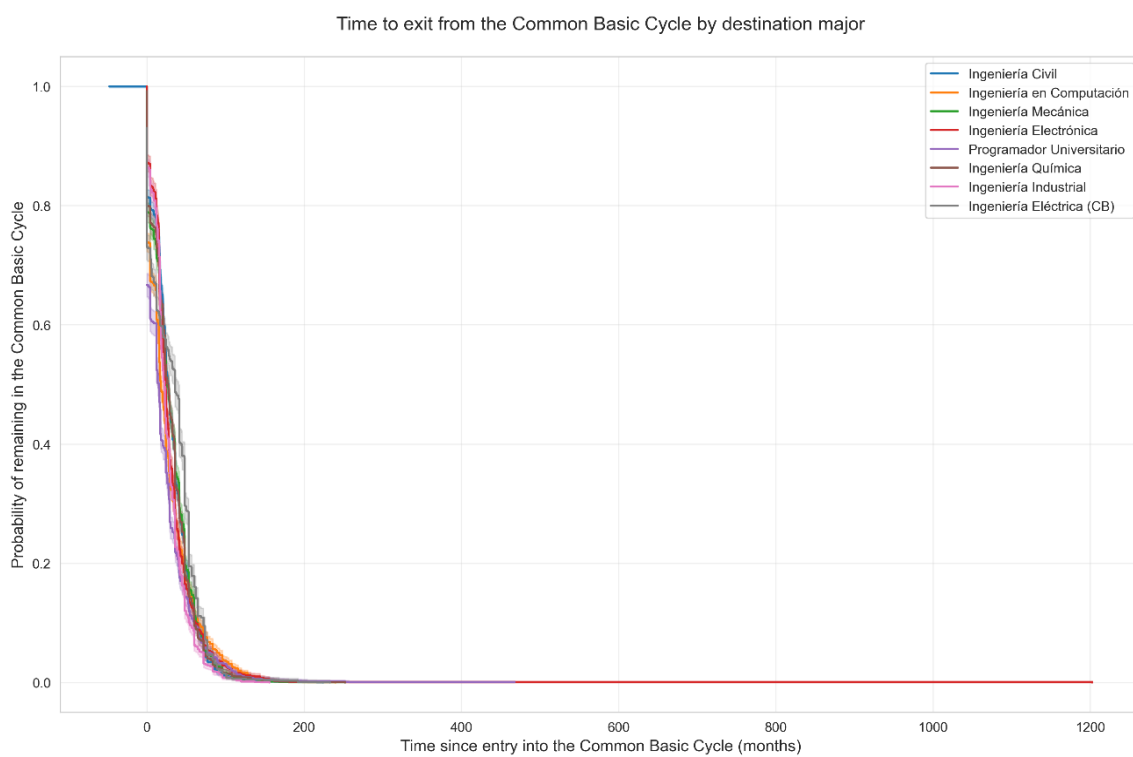

**3.4. Multi-major enrolment as adaptive exploration**

The extended dataset reveals that most students experience the CBC within a **single degree**. Approximately 85–90% of students hold only one major during the CBC (*n_majors_during_CBC = 1*), while about 10–15% enrol in **two or more majors** at some point during their CBC trajectory.

Outcome patterns differ sharply between these groups. Students with a single major have the highest CBC dropout rates, with roughly two-thirds leaving the system before reaching any upper-cycle subject. By contrast, students who explore **two majors** during the CBC have substantially lower dropout probabilities and higher rates of progression, either in their original or in an alternative degree. For those with three or more majors, dropout probabilities decrease further, although sample sizes are smaller.

These patterns suggest that multi-major enrolment during the CBC functions less as indecision and more as **adaptive exploration**: students who are willing or able to search for a better curricular fit are more likely to find a degree in which they can survive the CBC.

The analysis of **initial versus destination majors** provides a more detailed view. For traditional engineering degrees (for example, Civil, Industrial, Mechanical, Chemical, Electronic), most students who eventually reach the upper cycle start in the same degree. Multi-major trajectories are present but modest, and when they occur they tend to represent shifts within engineering (for example, from Mechanical to Industrial Engineering).

For shorter technical programmes, such as University Programmer, the picture is different. Students who **start** in these programmes exhibit very high CBC dropout rates (often exceeding three quarters of entrants), and only a small minority reach the upper cycle in the same degree. Contrary to common assumptions, relatively few of these students reappear as successful entrants in longer engineering degrees. In this dataset, shorter programmes act less as stepping-stones into engineering and more as **high-risk entry points** with limited progression. The analysis of enrolment flows reveals high stability in degree choice. Over 95% of students maintain their initial affiliation during the CBC. Cross-major flows are marginal and do not constitute a systemic pattern of vocational reorientation at this stage.

**3.5. Bottleneck and sentinel subjects in the Common Basic Cycle**

At subject level, the analysis identifies a small set of **structural bottlenecks** in the CBC. Introductory Mathematics modules – particularly *Introductory Calculus* and *Algebra and Analytic Geometry* – exhibit a combination of low pass rates and strong associations between failure and subsequent dropout. In the full sample, their global pass rates are often below 40%, and failing these subjects at least once is associated with a two- to three-fold increase in the hazard of leaving the system during the CBC.

These effects are not uniform across degrees. When pass rates are examined by destination major, the same Mathematics modules appear relatively permeable in traditional engineering degrees, where pass rates frequently exceed 80–90%, but remain extremely selective in shorter or more applied programmes, where pass rates can fall below 40%. Physics and introductory computing subjects show similar patterns, though typically with slightly higher pass rates and smaller hazard ratios. Figure 3 displays the odds ratios for dropout conditional on failing each CBC subject, stratified by destination major; cells with extreme values reflect combinations where failure is both rare and almost invariably terminal.

Figure 4 complements this risk perspective by displaying the underlying pass rates for key CBC subjects across destination majors. The heatmap reveals a striking pattern: traditional engineering programmes (Civil, Mechanical, Chemical, Electronic) exhibit pass rates consistently above 70% in most Mathematics and Physics modules, whereas shorter technical programmes such as University Programmer show pass rates below 50% in several foundational subjects. This asymmetry in baseline difficulty helps explain why the same CBC subjects function as navigable requirements for some students and as structural barriers for others

**Figure 3.** Odds ratios for dropout after subject failure by destination major. Each cell shows the odds ratio (OR) for dropping out of the CBC conditional on having failed the corresponding subject at least once. Colour scale represents log(OR): darker red indicates stronger association between failure and subsequent dropout. Empty cells denote subject–major combinations with insufficient data. Extreme values (e.g., OR > $10^6$) occur in combinations where failure is rare but nearly always followed by system exit.

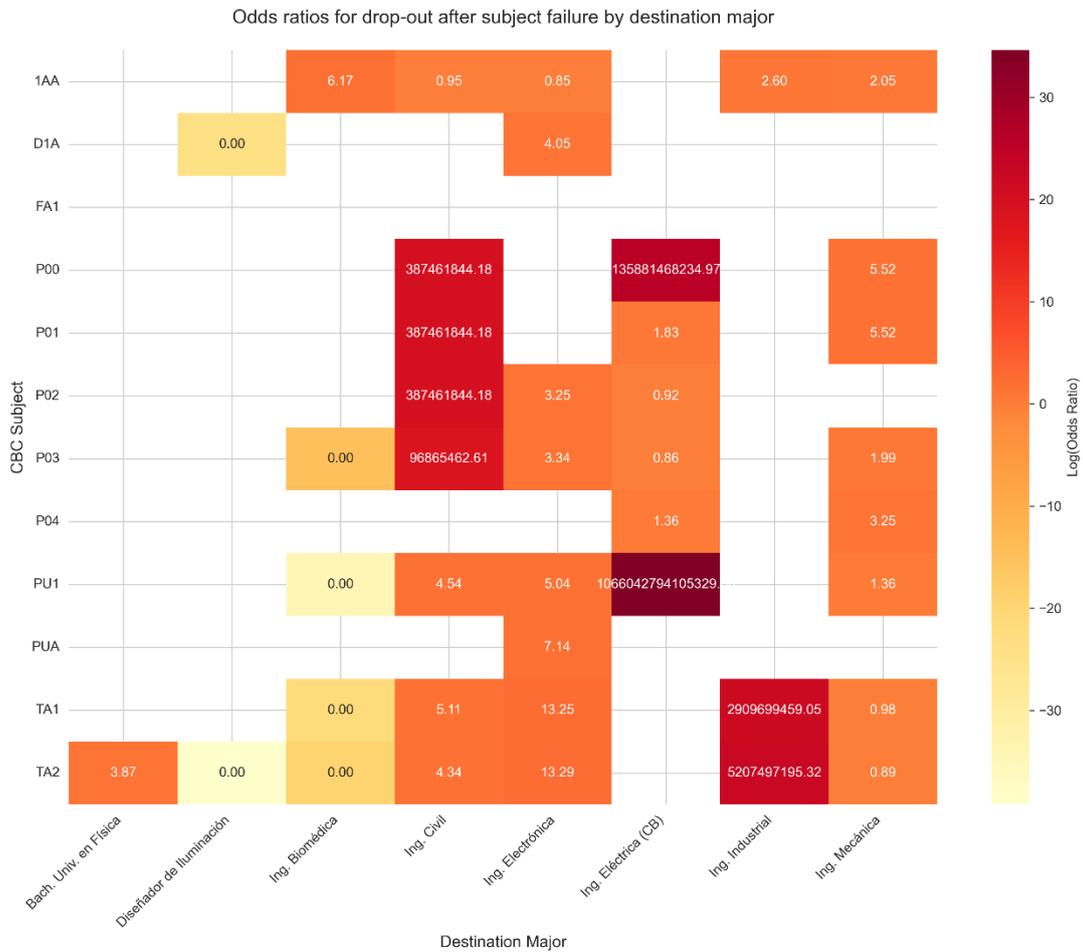

Figure **4.** Heatmap of pass rates in selected Mathematics Common Basic Cycle subjects by destination major (based on CBC_T02_subject_pass_rates_by_major)

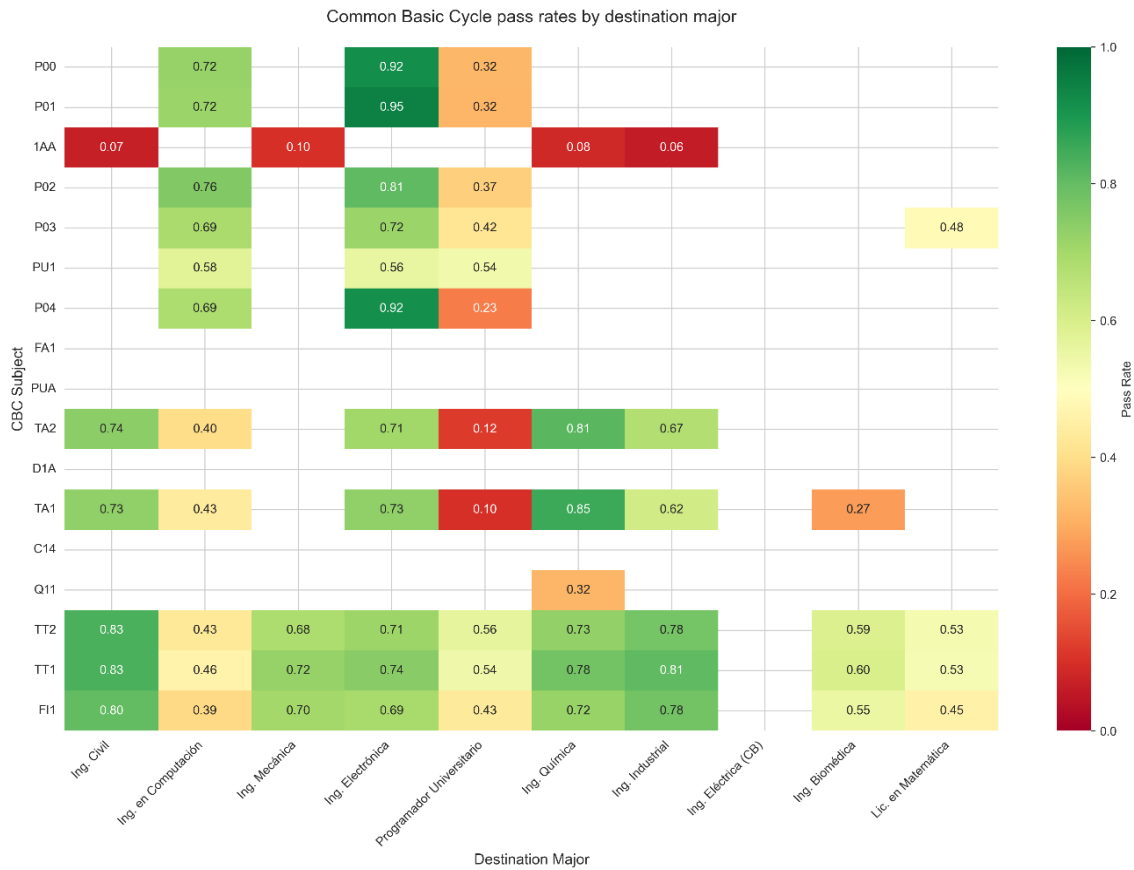

The **sentinel analysis**, which combines prevalence and adjusted odds ratios for dropout after failure at subject–major level, yields a small number of subject–major pairs meeting stringent criteria (prevalence of failure ≥ 0.10 and odds ratio ≥ 2.0 with p ≤ .01). These sentinel combinations tend to occur in smaller programmes where failure in a single key subject almost guarantees subsequent dropout. In larger engineering degrees, failure in Mathematics and Physics subjects is clearly associated with increased risk, but the diversity of trajectories and the presence of compensatory pathways make the statistical signals less extreme.

Taken together, these results suggest that the CBC is structured around a **concentrated set of mathematical and technical bottlenecks** whose effective strength varies by degree. In some programmes, these modules operate as demanding but navigable requirements; in others, they behave as narrow gates that few students manage to cross. The uneven distribution of pass rates and dropout risks across degrees reinforces the view of the CBC as a **differential sorting device**, rather than as a uniform levelling mechanism shared equally by all students.

## 4. DISCUSSION

The present study shows that a formally common basic cycle in exact sciences and engineering does not operate as a neutral bridge, but as a **differential filter** whose effects vary sharply across destination majors. This finding is consistent with classical models of student departure, which emphasise the interaction between student characteristics and institutional structures rather than individual deficits alone (Tinto, 1993; Yorke & Longden, 2004). In our case, the "structure" is not only the curriculum as a list of subjects, but the combination of **gateway mathematics courses**, **multi-major trajectories** and **programme-specific patterns of vulnerability** that emerge from longitudinal data.

### 4.1. CBC as a structured, non-neutral filter

At an aggregate level, the CBC behaves as a strong filter: a substantial fraction of students never reaches the upper cycle, even in a context where subjects are formally shared across majors (see Table 1. CBC outcomes by destination major). This is consistent with international evidence showing that early academic performance and first-year engagement are highly predictive of persistence and degree completion (Kuh et al., 2008; Paura & Arhipova, 2014; Singer & Willett, 2003). However, the key result here is that **the same CBC does not filter everyone in the same way**.

For some majors (for example, Physics or Chemical Engineering), CBC trajectories behave in the way one would expect from a "difficult but fair" first year: moderate failure rates, clear separation between successful and struggling students, and survival curves that diverge early and then stabilise (see Figure 2. Kaplan–Meier survival curves by destination major). For others, such as the University Programmer degree, the CBC looks more like a **structural dumping ground**: global failure rates are high and the odds of dropout are elevated for almost everyone, regardless of individual performance. This resonates with the notion of "weed-out" cultures in STEM, where early curricular segments function less as diagnostic stages and more as systematic exclusion mechanisms (Seymour & Hewitt, 1997; Seymour & Hunter, 2019).

### 4.2. Gateway mathematics as structural amplifiers, not merely "hard subjects"

The analysis of subject-level bottlenecks and sentinel subjects confirms the central role of **first-year mathematics** as a structural amplifier of risk. Algebra and Analysis appear as bottlenecks in several majors, but their consequences are not uniform. In some programmes, failing these subjects increases dropout risk modestly; in others, the same failure multiplies the odds of abandonment several times.

This pattern aligns with the international literature on **gateway mathematics courses** in STEM programmes. Studies on Calculus I in North American institutions have shown that early mathematics can act as a decisive branching point: students who struggle in these courses are disproportionately likely to leave STEM, even after controlling for prior preparation and aspirations (Ellis et al., 2016; Seymour & Hunter, 2019). Our results extend this insight in two ways. First, they show that **gateway effects are not properties of subjects in isolation**, but of **subject–major combinations**: the same algebra course functions as a mild hurdle for some majors and as a high-stakes gate for others. Second, they demonstrate that these effects can be quantified using survival and logistic models within a leakage-aware data layer (Paz, 2025; Therneau & Grambsch, 2000).

From a policy perspective, this matters because it shifts the narrative from "students cannot cope with mathematics" to "programmes are structurally fragile to specific mathematical failures". The same CBC subject may be pedagogically appropriate in itself, but its **position and coupling** within particular curricular paths turns it into a structural amplifier of dropout for some majors.

### 4.3. Multi-major trajectories: between risk and adaptive exploration

The extension of the pipeline to multi-major trajectories and post-2006 cohorts reveals that **simultaneous or sequential enrolment in several majors is both common and structurally relevant**. For a non-trivial proportion of students, CBC is not simply the first year of a single programme, but a **shared basin** from which trajectories branch into different destinations.

International research on major choice and person–environment fit suggests that students actively seek academic environments that match their interests and perceived strengths (Porter & Umbach, 2006; Seymour & Hunter, 2019). Our findings are compatible with this view but add a structural twist. In our data, multi-major trajectories play at least three distinct roles:

1. For some students, holding more than one major during the CBC appears as a **buffer**: it allows them to re-route from a highly selective engineering programme towards another major in exact sciences without leaving the institution.

2. For others, particularly in combinations that involve structurally fragile programmes, multi-major enrolment correlates with **diffuse trajectories** and increased dropout risk.

3. In a subset of cases, especially post-2006, multi-major paths are associated with **successful repositioning** from high-risk configurations (for example, initial enrolment as University Programmer) towards more stable engineering or science majors.

Figure 5 illustrates these student flow patterns from the CBC to final outcomes by destination major, showing the proportion of students who progressed to the upper cycle in the same major, switched to another major, or dropped out

**Figure 5.** Student flow from Common Basic Cycle to outcomes by destination major. Stacked bars show the proportion of students who progressed to the upper cycle in the same major (green), progressed to another major (blue), or dropped out (red). Programmes are ordered by progression rate in the same major.

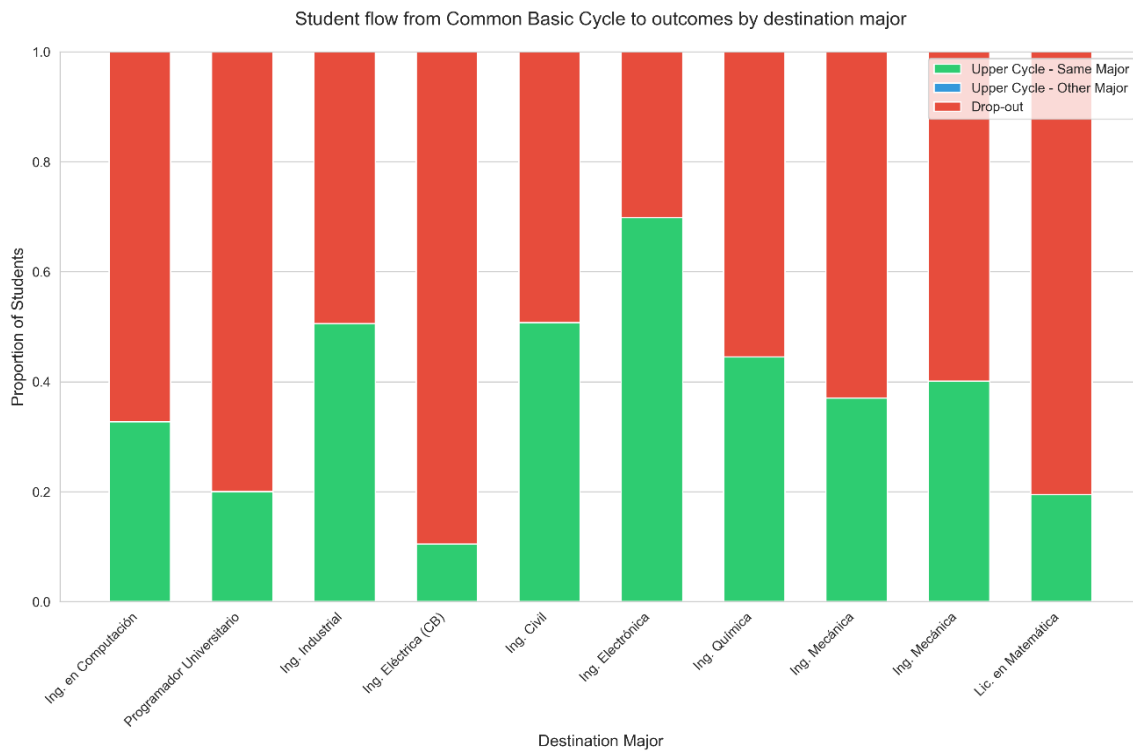

These patterns resonate with recent work on switching and relocation in STEM, which highlights that field changes are often rational responses to misalignment between students and programme cultures rather than simple loss of persistence (Seymour & Hunter, 2019). In this sense, the CBC functions both as a **risk structure** (when subject–major combinations amplify failure) and as an **opportunity structure** (when multi-major routes provide viable exits from structurally fragile configurations).

**4.4. Institutional context and temporal stability**

The decision to restrict some analyses to cohorts entering from 2006 onwards responds to a very specific institutional concern: the change of curriculum plans and CBC structure in the mid-2000s (Buchbinder, 2014; Universidad de Buenos Aires, n.d.). The fact that the main patterns — differential filtering by major, asymmetric gateway effects of mathematics, and the dual role of multi-major

trajectories — remain visible in the post-2006 subset strengthens the interpretation that we are observing **stable structural features** rather than artefacts of a particular plan or an exceptional cohort.

This is coherent with broader evidence that **first-year structures** retain their filtering role across reforms unless there are explicit interventions on teaching practices, assessment regimes or support structures (Kuh, 2008; Kuh et al., 2008; Yorke & Longden, 2004). In our case, the CBC appears to have preserved its basic logic as a common entry gate, but the internal distribution of risk has remained highly heterogeneous across majors.

### 4.5. Implications for CBC design and academic policy

Taken together, these findings support three main implications for CBC design in exact sciences and engineering:

1. **Move from global to targeted diagnosis.** Treating the CBC as a uniform "problem year" obscures the fact that risk is highly concentrated in specific **subject–major combinations**. Identifying sentinel subjects and structurally fragile trajectories allows institutions to focus support and monitoring where they are most needed, instead of applying generic first-year programmes (Paz, 2025; Singer & Willett, 2003).

2. **Re-think gateway mathematics as part of programme design, not only course difficulty.** In line with the calculus literature, our results suggest that simply "making mathematics easier" is unlikely to be sufficient (Ellis et al., 2016). What matters is how mathematics is positioned in each curriculum, what alternatives exist after an early failure, and whether students can re-route without incurring irreversible structural penalties.

3. **Recognise and manage multi-major trajectories as policy levers.** Multi-major enrolment is not a marginal anomaly but a central component of how students navigate the CBC. Institutions can either ignore this and treat switching as an individual problem, or explicitly design **structured internal pathways** that harness switching as a tool for improving person–environment fit (Porter & Umbach, 2006; Seymour & Hunter, 2019).

Finally, from a methodological perspective, this study illustrates the value of a **leakage-aware, multilevel data layer** for institutional decision-making (Paz, 2025). By combining survival models, logistic regression and explicit modelling of multi-major trajectories, it becomes possible to move beyond descriptive dropout rates and to articulate **programme-specific, structurally grounded narratives** about where and how the CBC filters students in exact sciences and engineering.

## 5. CONCLUSIONS AND POLICY IMPLICATIONS

This study examined a Common Basic Cycle (CBC) shared by exact-sciences and engineering degrees, asking whether it behaves as a neutral levelling device or as a differential filter across majors. Using longitudinal, leakage-aware data and explicitly modelling multi-major trajectories, the results point clearly in one direction: **the CBC is a strong, differential sorting mechanism**, not a homogeneous first-year experience.

Across cohorts, more than 60% of CBC entrants leave the system without ever reaching the upper cycle. After the 2005–2006 reform, this filtering becomes even more intense: progression to the upper cycle falls in most degrees, while dropout in the CBC rises. The formal promise of a "common" basic cycle hides the fact that, structurally, students are travelling through **very different corridors** depending on their destination programme. For some engineering degrees the CBC is demanding but permeable; for shorter and more applied degrees it is closer to an expulsion device, with only a small minority of entrants surviving to the upper cycle.

Subject-level analyses show that this filtering is organised around a small set of **structural bottlenecks**, particularly in Mathematics. Introductory Calculus and Algebra, together with core Physics and Computing modules, combine low pass rates with a substantial increase in the risk of leaving the system after failure. Crucially, the effective severity of these bottlenecks is not uniform: the same module can have pass rates above 80–90% in some programmes and below 40% in others, and its failure can be either a recoverable setback or a practical end-point. The CBC thus behaves as a system of **field-specific amplifiers**, where early mathematical challenges interact with the design and fragility of each programme.

At the same time, the analysis of multi-major trajectories complicates simple deficit narratives. Students who remain in a single major during the CBC are, on average, those with the highest dropout rates. In contrast, students who explore two or more majors have systematically lower dropout and higher progression. In this context, multi-major enrolment operates less as indecision and more as **adaptive exploration**: a way to search for a feasible trajectory within a structurally risky environment. Policies that restrict this exploratory behaviour risk concentrating failure among students trapped in poorly aligned initial choices.

Taken together, these findings have three broad implications for institutional policy. First, CBC design and evaluation should move beyond aggregate retention rates and examine **degree-specific filtering profiles**. It is not enough to ask whether the CBC "works" in general; institutions need to know for whom it works, in which programmes, and at what cost in terms of wasted time and opportunities.

Secondly, targeted interventions should focus on **structural bottlenecks rather than individual deficits**. This includes redesigning early Mathematics and Physics pathways, aligning assessment practices across programmes, and providing focused academic support precisely where the hazard of dropout spikes. Because the same subject plays different roles in different degrees, such interventions must be sensitive to local curriculum topologies, not only to course-level difficulty.

Thirdly, institutional regulations around **multi-major enrolment** in the CBC deserve reconsideration. The evidence presented here suggests that exploratory flexibility can function as a resilience mechanism, increasing the chances that students find a viable academic "home". Rather than discouraging multi-major trajectories, faculties may wish to make them more transparent and better supported, for example through explicit guidance on typical transition patterns, structured advising, and clear information on how the CBC connects to each degree.

The broader lesson is that a common basic cycle is never truly "common". Even when modules and syllabi are shared, the interaction between curriculum structure, assessment practices and student strategies produces **heterogeneous, path-dependent filtering** across engineering and exact-science programmes. Making these structural dynamics visible is a necessary step if faculties wish to move from implicit selection to deliberate design.

## REFERENCES

Buchbinder, P. (2014). *Reformas académicas y curriculares en la historia reciente de la Universidad de Buenos Aires: Una primera aproximación* (Documentos de discusión, N° 1). Universidad de Buenos Aires.

Ellis, J., Fosdick, B. K., & Rasmussen, C. (2016). Women 1.5 times more likely to leave STEM pipeline after calculus compared to men: Lack of mathematical confidence a potential culprit. *PLOS ONE, 11*(7), e0157447.

Hosmer, D. W., Lemeshow, S., & Sturdivant, R. X. (2013). *Applied logistic regression* (3rd ed.). Wiley.

Kuh, G. D. (2008). *High-impact educational practices: What they are, who has access to them, and why they matter*. Association of American Colleges and Universities.

Kuh, G. D., Cruce, T. M., Shoup, R., Kinzie, J., & Gonyea, R. M. (2008). Unmasking the effects of student engagement on first-year college grades and persistence. *Journal of Higher Education, 79*(5), 540–563.

Paura, L., & Arhipova, I. (2014). Cause analysis of students' dropout rate in higher education study programme. *Procedia – Social and Behavioral Sciences, 109*, 1282–1286.

Paz, H. R. (2025). A leakage-aware data layer for student analytics: The CAPIRE framework for multilevel trajectory modelling [Preprint]. National University of Tucumán.

Porter, S. R., & Umbach, P. D. (2006). College major choice: An analysis of person–environment fit. *Research in Higher Education, 47*(4), 429–449.

Seymour, E., & Hewitt, N. M. (1997). *Talking about leaving: Why undergraduates leave the sciences*. Westview Press.

Seymour, E., & Hunter, A.-B. (Eds.). (2019). *Talking about leaving revisited: Persistence, relocation, and loss in undergraduate STEM education*. Springer.

Singer, J. D., & Willett, J. B. (2003). *Applied longitudinal data analysis: Modelling change and event occurrence*. Oxford University Press.

Therneau, T. M., & Grambsch, P. M. (2000). *Modelling survival data: Extending the Cox model*. Springer.

Tinto, V. (1993). *Leaving college: Rethinking the causes and cures of student attrition* (2nd ed.). University of Chicago Press.

Universidad de Buenos Aires. (n.d.). *Recorrido histórico*. Secretaría de Asuntos Académicos.

Yorke, M., & Longden, B. (2004). *Retention and student success in higher education*. Open University Press.